\documentclass{article}
\usepackage{spconf,amsmath,graphicx}
\usepackage{subcaption}
\usepackage{booktabs}
\usepackage{multirow}
\usepackage{adjustbox}
\usepackage{graphicx}
\usepackage{lipsum}
\usepackage{verbatim}

\title{Fully unsupervised training of few-shot Keyword Spotting}
\name{Dongjune Lee$^{*}$, Minchan Kim$^{*}$, Sung Hwan Mun, Min Hyun Han, Nam Soo Kim}
\address{Department of Electrical and Computer Engineering and INMC, \\Seoul National University, Seoul, South Korea}
\makeatletter
\def\ps@IEEEtitlepagestyle{%
\def\@oddfoot{\mycopyrightnotice}%
\def\@evenfoot{}%
}
\def\mycopyrightnotice{%
{\footnotesize 978-1-6654-7189-3/22/\$31.00~\copyright~2023 IEEE\hfill} 
\gdef\mycopyrightnotice{}
}

\begin{document}
\ps@IEEEtitlepagestyle 
%
\maketitle
\def\thefootnote{*}\footnotetext{These authors contributed equally to this work.}
\def\thefootnote{\arabic{footnote}}

\begin{abstract}

For training a few-shot keyword spotting~(FS-KWS) model, a large labeled dataset containing massive target keywords has known to be essential to generalize to arbitrary target keywords with only a few enrollment samples. To alleviate the expensive data collection with labeling, in this paper, we propose a novel FS-KWS system trained only on synthetic data. The proposed system is based on metric learning enabling target keywords to be detected using distance metrics. Exploiting the speech synthesis model that generates speech with pseudo phonemes instead of texts, we easily obtain a large collection of multi-view samples with the same semantics. These samples are sufficient for training, considering metric learning does not intrinsically necessitate labeled data. All of the components in our framework do not require any supervision, making our method unsupervised. Experimental results on real datasets show our proposed method is competitive even without any labeled and real datasets.
\end{abstract}
%

\section{Introduction}
\label{sec:intro}
Keyword spotting~(KWS) is to identify a target keyword in the continuous audio streams, which is broadly deployed as a front door to voice assistants in several edge devices such as smartphones and AI speakers. In general, KWS systems predetermine target keywords and are directly optimized for selected keywords. Although existing predefined KWS models show high detection performance~\cite{rybakov2020streaming, berg2021keyword, kim2021broadcasted}, the necessity of a large dataset containing target keywords and inflexibility of changing target keywords hinder KWS models from expanding to various applications. When it comes to user-defined KWS, users can customize the target keywords with only a few enrollment samples~\cite{chen2018meta, mazumder2021few, huh2021metric, parnami2022few} or in the form of string~\cite{sacchi2019open, shin2022learning}. 
Few-shot KWS~(FS-KWS) especially has shown its feasibility through meta learning~\cite{chen2018meta}, transfer learning~\cite{mazumder2021few}, and metric learning~\cite{huh2021metric, parnami2022few}, operating on the few-shot detection scenario. 
These approaches typically require learning from a large corpus with lots of different keywords to secure generalization on unseen keywords with few samples. However, user-defined KWS despite its potential in diverse usability has been underexplored due to the deficiency of a large high quality public corpus.

Recently, with dramatic advances in deep generative models, there have been several approaches that leverage generative models as a data source to compensate for insufficient or unavailable labeled data. For example, data augmentation using synthesized data has been explored in KWS and automatic speech recognition~(ASR)\cite{lin2020training, chen19f_interspeech, hu2022synt++}. Text-to-speech~(TTS) models are utilized to supplement the less frequent utterances such as named entities which are difficult to collect in the real world. In the vision domain, Besnier et al.~\cite{besnier2020dataset} effectively trains a classifier with several learning strategies only using synthetic image data generated by a conditional GAN. Moreover, Jahanian et al.~\cite{jahanian2021generative} used an unconditional GAN as a data source for representation learning. In ~\cite{jahanian2021generative}, searching in latent space of GAN offers multi-view data sharing the semantics which is necessary for training contrastive objective of representation learning. The aforementioned approaches show that the synthesized dataset 
has the potential to substitute a real dataset, not only bounded to taking a role as a subsidiary dataset.

In this paper, we propose a novel framework for FS-KWS trained on only synthetic data. The proposed framework is based on metric learning so as to detect target keywords with few-shot enrollments. We assume that a good FS-KWS model can extract phonetic information from any short utterances and make clusters of any arbitrary phonetic chunks from various voices. Our motivation stems from the question of whether a labeled KWS dataset is indispensable in metric learning if we can acquire a large bunch of utterances with the same pronunciation by a different route. The notable point is that the metric learning objective in FS-KWS does not require any explicit textual supervision for training. Instead, we exploit pseudo-TTS model~\cite{kim2022transfer}, which is trained on a large-scale unlabeled speech corpus. The pseudo-TTS model takes pseudo phoneme sequence extracted from wav2vec2.0~\cite{baevski2020wav2vec} and reference speech as inputs, and returns utterances with various speakers and prosody reflected. Using a phonetic representation of wav2vec2.0, the pseudo-TTS system takes the role of decomposing the utterances to the fine-grained factors and reassembling to the suitable form for metric learning. We notice that our proposed method is trained in a fully unsupervised manner as all of the components including wav2vec2.0 and pseudo-TTS are trained without any supervision. From our experiments, we find that even without using any real and labeled data, the proposed model demonstrates high performance on KWS datasets~\cite{baevski2020wav2vec, warden2018speech}. The experimental results show the potential of utilizing unsupervised speech synthesis for user-defined KWS.

\section{Backgrounds}
\label{sec:format}
\subsection{Metric Learning based KWS}
Most current KWS models are trained and evaluated under classification objectives, treating all the possible non-target keywords as a single class. Although there exist countless non-target sounds, limited non-target samples are used during training, which can inhibit the detection performance. Furthermore, customizing target keywords in the classification scenario is difficult since the models are not trained to distinguish the diversity of non-target sounds and only respond to the original target keywords. To overcome these problems, several metric learning based KWS models have recently arisen. The goal of metric learning for KWS is to acquire a general representation for KWS and detect target keywords using distance metrics. For example, Huh et al.\cite{huh2021metric} explore several metric learning objectives such as triplet loss\cite{hoffer2015deep} and prototypical loss\cite{snell2017prototypical} for training KWS. In addition, Kim et al.\cite{kim2022dummy} suggest a multiple dummy prototype generator to handle open-set queries efficiently. Considering that KWS is closer to the detection task rather than the classification task in the real-world scenario, metric learning methods are advantageous over classification approaches, effectively tackling unknown category samples via distance metrics. 

\subsection{TTS with pseudo phoneme}
In \cite{kim2022transfer}, TTS with pseudo phoneme was firstly proposed for the transfer learning framework for low-resource TTS. The pseudo phoneme is a phonetic token that can be obtained without any text labels. Instead, the pseudo phoneme is extracted from an unlabeled speech by k-means clustering of wav2vec2.0 embeddings which contain rich phonetic information. Kim et al.~\cite{kim2022transfer} pre-trained VITS~\cite{kim2021conditional} based TTS model using the pseudo phoneme and fine-tuned the pre-trained model on a small amount of transcribed corpus with real phoneme. For convenience, we refer to the pre-trained model in \cite{kim2022transfer} to pseudo-TTS model. The pseudo-TTS model takes pseudo phoneme sequence and a reference speech as inputs and returns synthesized speech that contains phonetic information of pseudo phoneme with the speaker and prosody of reference speech. Using the pseudo-TTS model, with only unlabeled speech corpus, we can get massive groups of the same pronunciation spoken by various speakers with diverse prosody.

\section{Proposed method}
\label{sec:pagestyle}

\begin{figure}[ht]
 \centering
 \begin{subfigure}{\columnwidth}
 \includegraphics[width=\columnwidth]{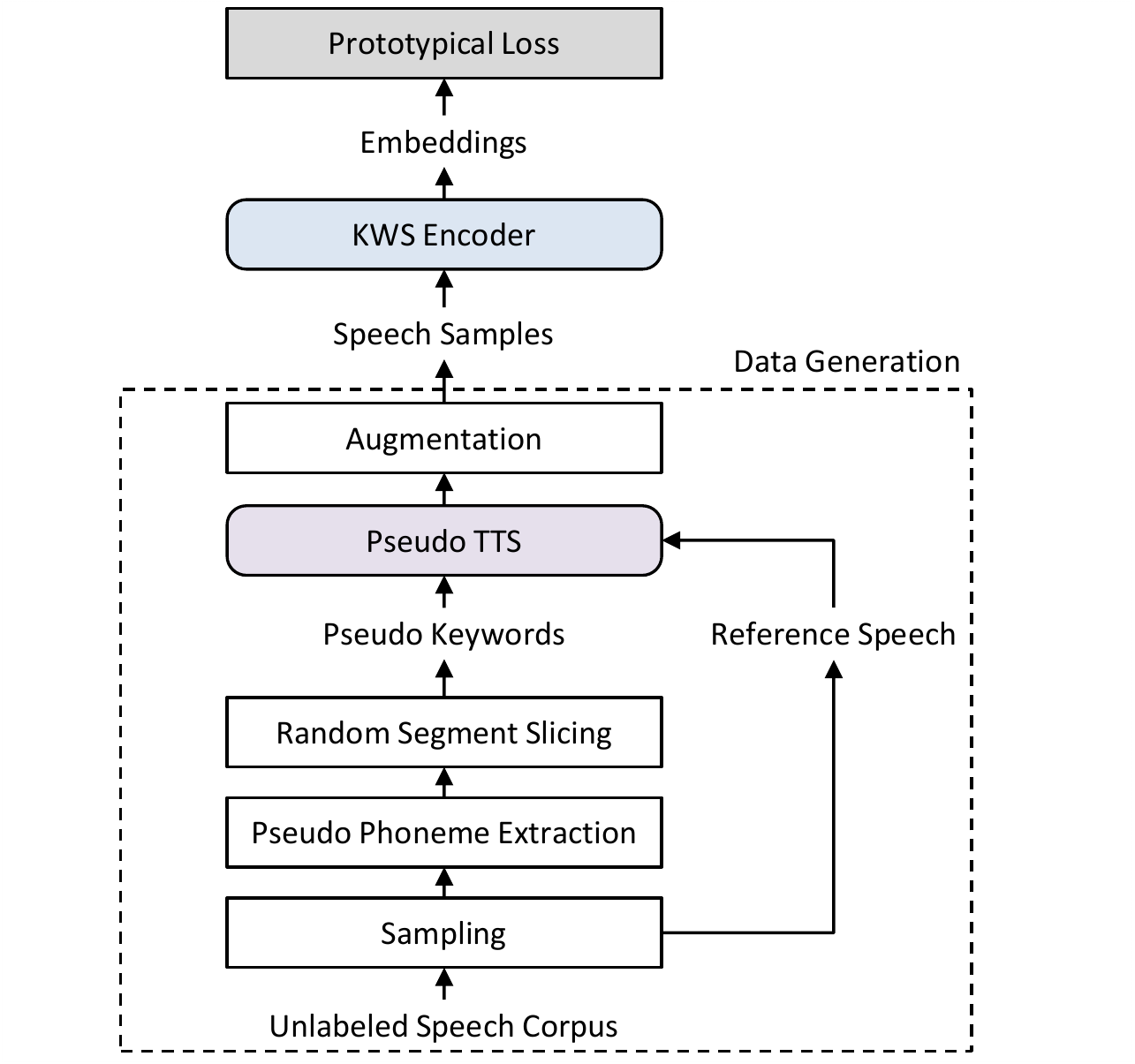}
 \caption{Training procedure}
 \label{fig:speech_production}
 \end{subfigure}\vspace{0.2cm}
 \begin{subfigure}{\columnwidth}
 \includegraphics[width=\columnwidth]{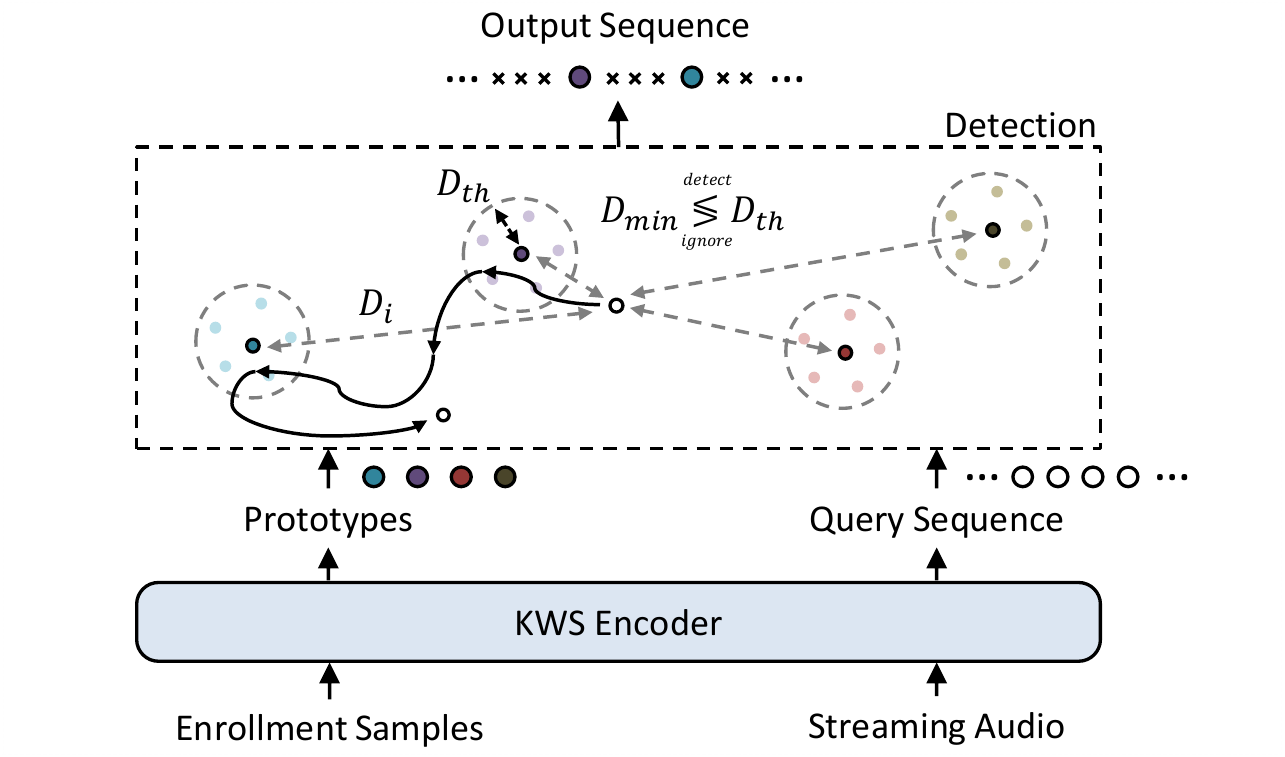}
 \caption{Inference procedure}
 \label{fig:speech_production}
 \end{subfigure}
 \caption{The entire framework of the proposed method} 
 \label{figure1} 
 \end{figure}

In this section, we describe the overall framework of the proposed method. The proposed method uses the pseudo-TTS model as a data source for training FS-KWS with metric learning objective. The entire framework is depicted in Figure\ref{figure1}.

\subsection{Training}
\label{ssec:subhead}

\subsubsection{Training objective}
\label{ssec:objective}

Among the methods of metric learning, we adopt prototypical networks~\cite{snell2017prototypical} which operate in $N$-way $K$-shot classification, where $N$ and $K$ denote the number of classes and supports respectively. At each iteration, the encoder takes $N\times(K+1)$ input speech $x_{1:N, 1:K+1}$ and returns output embeddings. Here, we fix the number of queries per class to 1 for simplicity, so that $x_{n, 1:K}$ represents the support set and $x_{n, K+1}$ represents the query for class $n$. The training objective is formulated as (1)-(3).

\begin{equation} \label{eq:kld1}
  c_n = \frac{1}{K} \sum_{k = 1}^{K} f_\phi(x_{n,k}),
\end{equation}
\begin{equation} \label{eq:kld2}
  p(y=n|x) = \frac{\exp(-dist(f_\phi(x),c_n))}{\sum_{n'=1}^{N}{\exp(-dist(f_\phi(x),c_{n'}}))},
\end{equation}
\begin{equation} \label{eq:kld3}
  L = -\frac{1}{N} \sum_{n=1}^{N} \log{p(y=n|x_{n,K+1})}.
\end{equation}

In (1), $c_n$ is a prototype of class $n$ and $f_\phi$ denotes the encoder parameterized by $\phi$. In (2), $dist(\cdot,\cdot)$ can be any distance metric for which we used euclidean distance. The objective $L$ represents the cross-entropy loss for the given few-shot classification.

\subsubsection{Data generation using pseudo-TTS}
\label{ssec:subhead}

In the proposed method, the entire data is generated by the pseudo-TTS model~\cite{kim2022transfer}. Although the pseudo-TTS model was originally designed for the pre-training stage of transfer learning, we exploit it to generate arbitrary speech under various conditions. The data generation is processed as follows. First, we sample speech from an unlabeled speech corpus and extract pseudo phonemes. Then, we randomly crop the pseudo phoneme sequences to get the arbitrary pseudo keywords with lengths sampled from $(L_{min}, L_{max})$. Both $L_{min}$ and $L_{max}$ are hyperparameters of data generation. After, we sample reference speech from a speech corpus for the speaker and prosody variability. In consequence, we can generate unlimited amounts of pronunciation chunks in an unsupervised manner.

Although the multi-view samples of each pseudo keyword can be obtained by the pseudo-TTS model, the domain mismatch between synthesized and real audio can significantly harm the performance. To make the generated speech more realistic, we used several augmentation schemes. First, we scale the volume of each speech by adjusting the maximum value to a randomly sampled value. Second, we add reverberation by convolution with simulated RIR corpus. Lastly, we add noise to the generated samples using additional noise corpus with various SNR.

\subsubsection{Buffered samples for efficient training}
\label{ssec:subhead}
Generating thousands of seconds of speech demands excessive computations and memory consumption, which induces various practical issues such as slow training. Instead of generating $N\times(K+1)$ samples every iteration, we use a data buffer for efficient training. At the beginning of training, we initialize the buffer with $M_{buffer}\times(K+1)$ samples, where $M_{buffer}> N$. At every iteration, we randomly select $N$ classes from the buffer for training and enqueue and dequeue samples of $M_{update}$ classes in the buffer. In the case of $M_{update}\ll N\ll  M_{buffer}$, we can save a considerable amount of speech at every iteration without overfitting to the current buffer.

\subsection{Inference}
\label{ssec:subhead}

First, we collect utterances of target keywords for enrollment. The prototype $c_n$ of each class $n$ is calculated in the same manner as (1). Then, the inference process follows the steps in (4)-(5).
\begin{equation} \label{eq: kld4}
  \hat{y} = \underset{n}{\mathrm{argmin}}\ dist(f_\phi (x), c_n),
\end{equation}
\begin{equation} \label{eq: kld4}
  y_{pred} = 
  \begin{cases}
    \hat{y}, & \mbox{if}\ dist(f_\phi (x), c_{\hat{y}}) < D_{th}  \\
    \hfil N+1, & \hfil \mbox{else}
  \end{cases}.
\end{equation}
In (4)-(5), $\hat{y}$ denotes the candidate class with the maximum ~$p(y=n|x)$, where $x$ denotes input query speech and $y$ denotes the class label. Unlike the training objective which conducts classification, the whole inference process serves as a detection task. Thus, the final output class is determined by the distance to the candidate class. Here, $N+1$ means the `unknown' class. The detection threshold $D_{th}$ can be tuned for the desired false positive rate.

\begin{table*}[h]
\centering
\caption{Results of the experiments. All of the measures are tried 100 times, and the average scores of accuracy and AUROC are written in (\%) with 95\% confidence intervals.}
\label{tab:my-table}
\begin{tabular}{@{}c|c|ccc|ccc@{}}
\toprule
\multirow{2}{*}{\textbf{Method}} & \multirow{2}{*}{\begin{tabular}[c]{@{}c@{}}\# of\\ supports\end{tabular}} & \multicolumn{3}{c|}{\textbf{GSCv1}}                                      & \multicolumn{3}{c}{\textbf{MSWC}}                                        \\
                                 &                                                                           & Acc (target)           & Acc (total)            & AUROC                  & Acc (target)           & Acc (total)            & AUROC                  \\ \midrule
Supervised (MSWC)                & \multirow{3}{*}{1-shot}                                                   & 69.0$\pm$1.67          & 66.0$\pm$1.03          & 77.8$\pm$1.03          & \textbf{72.8$\pm$1.67} & \textbf{64.8$\pm$1.31} & \textbf{75.9$\pm$1.45} \\
Unsupervised (w/o aug)           &                                                                           & 49.5$\pm$1.21          & 51.4$\pm$0.64          & 64.2$\pm$0.83          & 54.0$\pm$1.36          & 54.2$\pm$0.73          & 66.9$\pm$0.90          \\
Unsupervised (w aug)             &                                                                           & \textbf{76.0$\pm$1.29} & \textbf{69.6$\pm$0.75} & \textbf{81.1$\pm$0.83} & 67.1$\pm$1.20          & 62.2$\pm$0.68          & 74.5$\pm$0.87          \\ \midrule
Supervised (MSWC)                & \multirow{3}{*}{5-shot}                                                   & 90.5$\pm$0.53 & 80.6$\pm$0.44          & 89.4$\pm$0.38          & \textbf{91.7$\pm$0.66} & \textbf{81.8$\pm$0.82} & \textbf{90.9$\pm$0.75} \\
Unsupervised (w/o aug)           &                                                                           & 78.0$\pm$0.82          & 64.3$\pm$0.48          & 74.4$\pm$0.57          & 78.3$\pm$1.01          & 66.9$\pm$0.61          & 77.7$\pm$0.69          \\
Unsupervised (w aug)             &                                                                           & \textbf{91.0$\pm$0.45} & \textbf{82.0$\pm$0.39} & \textbf{90.8$\pm$0.36} & 86.5$\pm$0.79          & 76.2$\pm$0.56          & 86.7$\pm$0.58          \\ \midrule
Supervised (MSWC)                & \multirow{3}{*}{20-shot}                                                  & \textbf{93.7$\pm$0.26} & 83.8$\pm$0.33          & 91.5$\pm$0.30          & \textbf{94.4$\pm$0.40} & \textbf{85.6$\pm$0.57} & \textbf{93.5$\pm$0.47} \\
Unsupervised (w/o aug)           &                                                                           & 86.8$\pm$0.46          & 67.9$\pm$0.50          & 76.7$\pm$0.59          & 86.1$\pm$0.62          & 70.5$\pm$0.78          & 80.4$\pm$0.91          \\
Unsupervised (w aug)             &                                                                           & 93.3$\pm$0.31          & \textbf{85.0$\pm$0.33} & \textbf{92.9$\pm$0.28} & 91.4$\pm$0.43          & 80.3$\pm$0.42          & 89.7$\pm$0.40          \\ \bottomrule
\end{tabular}
\end{table*}

\section{Experiments}
\label{sec:typestyle}

\subsection{Experimental Settings}
\label{ssec:subhead}

\subsubsection{Pseudo-TTS implementation}
\label{ssec:subhead}
To implement the pseudo-TTS model, we used an English part of Multilingual LibriSpeech~(MLS)~\cite{pratap2020mls}, one of the largest public speech corpora with high quality. We did not use any transcript of MLS to satisfy the unsupervised condition. For pseudo-phoneme extraction, we leveraged the pre-trained wav2vec2.0 model from an open source\footnote{https://huggingface.co/facebook/wav2vec2-large-xlsr-53}. We mostly followed the implementation details of \cite{kim2022transfer} including model architecture, hyperparameters, and pseudo phoneme extraction. The only difference is the training dataset and the sampling rate, which was changed from 22kHz to 16kHz. We trained the pseudo-TTS model for 500k iterations. When generating speech, we also sampled pseudo keywords from the English part of MLS.

\subsubsection{Model Architecture}
\label{ssec:subhead}
We adopt TC-Resnet~\cite{choi19_interspeech} with several modifications for the encoder. TC-Resnet is originally composed of residual 1D convolution and batch-normalization blocks, followed by an average pooling layer to aggregate along the time axis. We replace an average pooling layer with a single GRU layer to efficiently capture the temporal dynamics. Then, the output embedding corresponds to the linear projection of the last frame of the GRU layer. We also set the scale factor k of TC-Resnet to 2 for higher model complexity. The model takes 40-dimensional MFCC as an input sequence and returns an output embedding with 192 dimensions.

\subsubsection{Implementation Details}
\label{ssec:subhead}

We trained the KWS model on a 512-way 5-shot scenario. For the data buffer described in 3.1.3, we set $M_{buffer}$ to 32768, and $M_{update}$ to 1 for our experiments. For pseudo keyword generation, we set $L_{min}$ to 10 and $L_{max}$ to 20. These lengths were determined to generate plausible spoken keywords. For data augmentation, we used noise corpus from \cite{snyder2015musan} and RIR coefficients from \cite{ko2017study}. The maximum value for volume scale is sampled from $(0.2, 0.9)$ and SNR is sampled from $(10, 20)$. Noise injection and reverberation are applied with the probability of $0.9$. We trained the model with Adam optimizer with the learning rate $0.001$ and cosine scheduler for 300k iterations.

\subsubsection{Evaluation}
\label{ssec:subhead}
To verify the operation of our proposed method in real-world applications, we evaluated the proposed method on two real KWS datasets: Google Speech Commands version one~(GSCv1)~\cite{warden2018speech} and Multilingual Spoken Word Corpus~(MSWC)~\cite{mazumder2021multilingual}. We measured top-1 accuracy and Area Under ROC~(AUROC) for the experiments, which will be described in detail below.\\
\textbf{GSCv1}:
GSCv1 is the most common KWS corpus of 30 keywords in total. We followed the official division of train, validation, and test set of ~\cite{warden2018speech}, and did not use the subset of background noise for our experiments. Since the proposed method operates on the open-set and few-shot scenario, we randomly selected target and unknown keywords. To be specific, we sampled 10 target keywords and treated the remaining 20 keywords as the unknown keyword. Then, we sampled $K$ supports from the train set of each target keyword for enrollment and evaluated the test division. We tested for $K=1, 5, 20$ respectively.\\
\textbf{MSWC}:
MSWC is the largest public KWS corpus of multiple languages. Among them, we used the English part of MSWC~(MSWC-en). MSWC-en consists of about 7 million samples of 39 thousand keywords. Firstly, we sampled 100 test keywords that have a number of samples in (2000, 3000). For each test keyword, the first 250 samples are used for the test, and the rest samples are regarded as the pool of supports. Then, at every evaluation, we sampled 10 target keywords and 20 unknown keywords from the test keywords to match the experiment condition with GSCv1. Finally, we randomly enrolled $K$ samples from the support pools and tested them on the selected test set, where we set $K=1, 5, 20$.\\
\textbf{Baseline}:
For comparison, we implemented three models: Unsupervised~(w aug), Unsupervised~(w/o aug), and Supervised~(MSWC). Unsupervised~(w aug) is the proposed model with all of the described augmentation methods. Unsupervised~(w/o aug) is trained on the entirely same condition of Unsupervised~(w aug), but without any augmentation. To show the effectiveness of unsupervised training, we additionally set a model trained with supervision. Supervised~(MSWC) is trained on MSWC-en, except for the test keywords under the assumption that the size of MSWC-en is large enough for training FS-KWS. For training Supervised~(MSWC), we sampled $N$ classes and $K$ utterances from each class every iteration as we did for the proposed method. All of the other conditions remained the same, excluding the training dataset. We notice that data augmentation was not used for training Supervised~(MSWC).\\
\textbf{Measure}:
We used top-1 accuracy and AUROC for measurement. The accuracy is divided into Acc~(target) and Acc~(total). Acc~(target) is measured only for the target keywords~(10 classes) to show the classification performance. On the contrary, Acc~(total) is the accuracy of classification including the unknown class~(11 classes). As Acc~(total) is determined by $D_{th}$, we fixed $D_{th}$ to the threshold at equal error rate~(EER) of the binary classification of whether the sample is unknown. AUROC is also calculated for the same binary classification. Since the measures are susceptible to not only the selected classes but also selected supports, we tested 100 times and reported the average score with 95\% confidence intervals.\\

\begin{figure}[]
 \centering
 \includegraphics[width=\columnwidth]{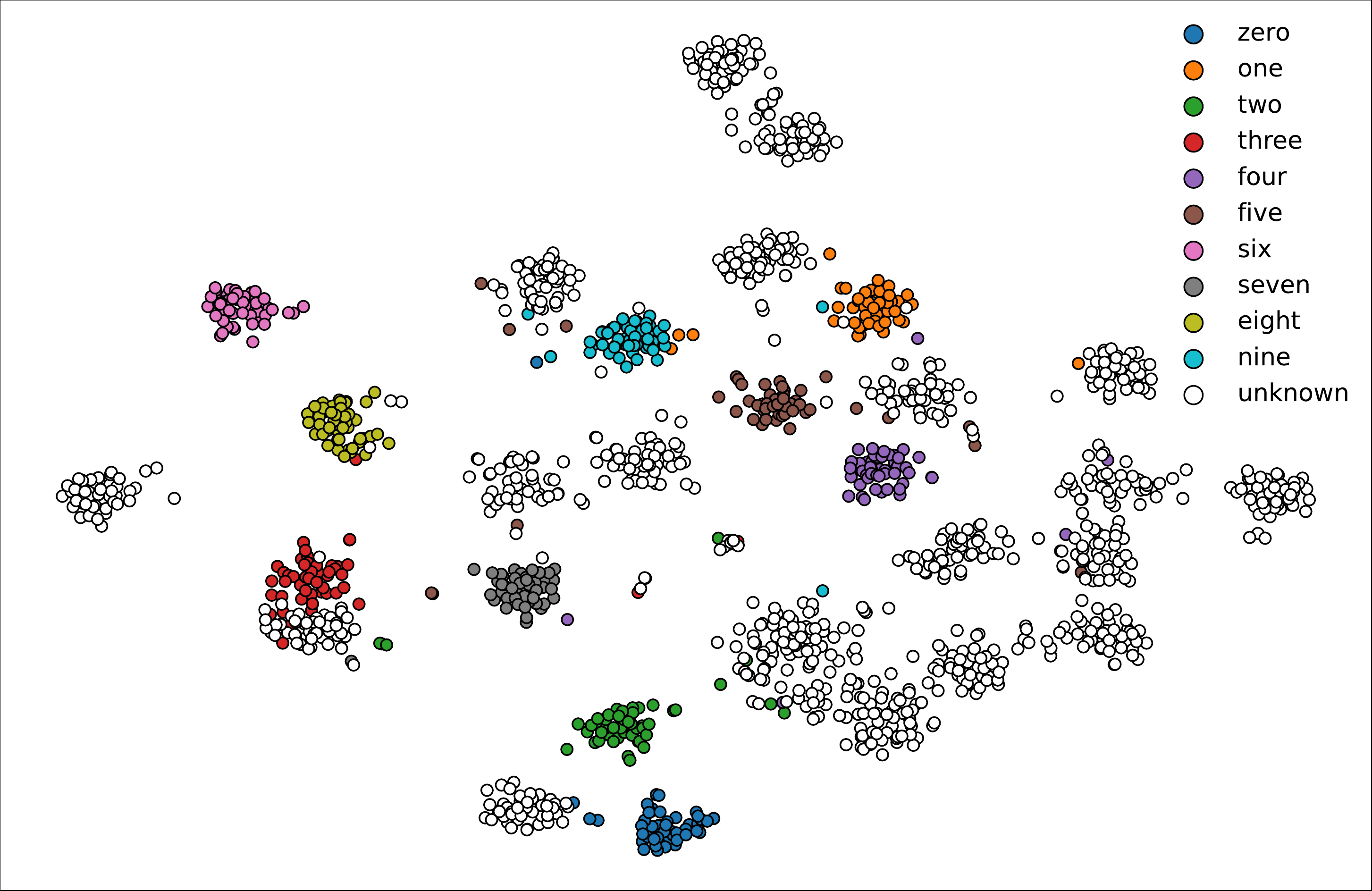}
 \caption{Visualization of embeddings of GSCv1 using t-SNE. Ten keywords~(digits) are treated as target keywords, and the remaining 20 keywords are labeled as `unknown'. We plotted 50 randomly selected samples of each keyword.}
 \label{fig:t-SNE}
 \end{figure}

\subsection{Results}
\label{ssec:subhead}
The overall results are presented in Table 1. First of all, the trend of both accuracy and AUROC measures in all the cases showed almost the same behavior, pointing out that classification performance goes along with detection performance. In addition, all the scores increased as more supports were provided. Also, the variance of scores had a tendency to be larger as the number of supports was smaller. These two tendencies imply the necessity of a sufficient number of supports for enrollment. The most notable point in the results is the performance gap between the model with augmentation and the model without augmentation. Showing significant margins in all the scores, the results imply that generating clean speech itself does not always guarantee the utmost performance in the downstream tasks. However, at the same time, even simple data augmentation schemes can considerably mitigate the domain mismatch between the synthesized speech and the real speech. When comparing the performance on the real dataset between the proposed method and baseline model, different consequences were shown in GSCv1 and MSWC. In GSCv1, the proposed method with augmentation mostly showed slightly higher scores, but showed the opposite in MSWC. This result, we believe is mainly due to a mismatch between train and test domain. More importantly, however, the fact that the proposed model showed higher performance in GSCv1 without any labeled and real dataset for training bolsters our contributions.

For a subjective analysis, we visualized the embedding space of the proposed model in Figure 2. We plotted embeddings from GSCv1, where we colorized ten keywords~(digits) and assigned the remaining keywords as `unknown'. Figure 2 shows that embeddings from the same class form cluster well so that the discriminative nature of each cluster verifies the validity of both the proposed training and inference scheme.

\section{Conclusion and Discussion}
\label{sec:print}

In this work, we propose an unsupervised training scheme for few-shot KWS by leveraging a speech synthesis system named pseudo-TTS. The proposed method is based on prototypical networks trained on a massive bunch of multi-view samples with the same phonetic information. For this requirement, we used the pseudo-TTS model which uses pseudo phonemes extracted in an unsupervised manner from wav2vec2.0 representation. The experimental results present that the proposed KWS system shows comparable performance to the supervised training in the real speech domain.

From this work, we show that the recent speech synthesis systems with high fidelity and controllability have the potential to replace the labeled dataset for user-defined KWS. However, as there are a lot of approaches that haven't been explored yet, we can extend this work in various directions for future works. First of all, we assume that the diversity of synthesized speech is the most important key to improvement. Although recent speech synthesis systems can effectively control factors including speaker, speech rate, and prosody, there remains a limitation in controllability and a gap of variability between real speech recorded from various conditions. Thus, we will follow up the research on the controllability of speech synthesis and adopt them for keyword generation. In addition, domain adaptation or augmentation can be a breakthrough in utilizing synthetic data, boosting model robustness. Beyond the augmentation methods tried in this work, carefully designed domain matching algorithms should be more explored. Lastly, general improvement in architecture designs and manipulating other metric learning methods can lead to better performance.

\section{Acknowledgements}
\label{sec:print}

This work was supported by Institute of Information \& communications Technology Planning \& Evaluation (IITP) grant funded by the Korea government(MSIT) (No.2021-0-00456, Development of Ultra-high Speech Quality  Technology for Remote Multi-speaker Conference System).

\bibliographystyle{IEEEbib}
\bibliography{strings,refs}

\end{document}